\documentclass[%
reprint,
 amsmath,amssymb,
 aps,
pra,
]{revtex4-2}

\usepackage{graphicx}
\usepackage{dcolumn}
\usepackage{bm}
 \usepackage{hyperref}
\begin{document}

\preprint{APS/123-QED}


\title{Single-photon phase spectrum recovery from the Hong-Ou-Mandel dip}

\author{Yuhang Lei}
\email{leiyuhang@tju.edu.cn}
\author{Wen Zhao}
\author{Liang Cui}
\email{lcui@tju.edu.cn}
\author{Xiaoying Li}
\affiliation{%
College of Precision Instrument and Opto-Electronics Engineering, Key Laboratory of Opto-Electronics Information Technology, Ministry of Education, Tianjin University, Tianjin 300072, People’s Republic of China
}%

\date{\today}

\begin{abstract}
 Characterizing the temporal-spectral profile of single photons is essential for quantum information protocol utilizing temporal mode for encoding. Based on the phase retrieval algorithm, we present a method to reconstruct the phase spectrum difference between two wave packets from their Hong-Ou-Mandel dip, and intensity spectra. Our confirmatory experiment with weak coherent wave packets demonstrated the accuracy of the reconstructed phase spectrum difference to within $\pm$ 0.1 rad. This method is generalizable to the measurement of unknown single-photon wave packets with the aid of a reference wave packet, requiring only the collection of one-dimensional data, which simplifies and expedites the process.
\end{abstract}

\maketitle

\section{Introduction}
The mode profile of single photons fully determines their characteristics and interference phenomena, enabling quantum information protocols to encode information across different degrees of freedom   \cite{cod_1-PhysRevLett.81.3563,cod_2-PhysRevA.87.062322,cod_3-capability3-Nunn:13,cod_4-PhysRevLett.112.133602,cod_5-PhysRevX.5.041017,cod_6-Schwarz_2016}. The temporal-spectral degree of freedom supports high-dimensional encoding frameworks   \cite{highdimension_Thiel_2017}, 
 which can enhance the capability of information encoding   \cite{capability2-Zhong_2015,cod_5-PhysRevX.5.041017}.
This feature necessitates the characterization of the temporal profile of single photons, which can be represented by the intensity spectrum and the phase spectrum.
While the intensity spectrum has mature measurements at the single-photon level   \cite{Smith2009PhotonPG_MONO_jsi,Davis2016PulsedSS_CHIRPED_JSI}, the phase spectrum, is harder to characterize. The phase spectrum measurement usually requires nonlinear-based techniques,
such as frequency-resolved optical gating  \cite{Trebino1993UsingPR_Trebino} and and phase spectrum interferometry for direct electric field reconstruction \cite{SPIDER-Iaconis:98}, which, however, can not be extended to the single-photon regime directly  to the weak nonlinear effects \cite{PhysRevLett.121.083602_shearing_singlephoton}. 
Recently, measurment of the temporal-spectral profile of single photons has been demonstrated through
nonlinear methods that project them to the orthogonal temporal modes  \cite{ansari2018tomography_DAVIS23,PhysRevLett.124.213603_Nanshijie}, or using linear interferometry  \cite{wasilewski2007spectral_DAVIS20,Thiel:19_FrqRslvHOM_singlephoton,PhysRevLett.121.083602_shearing_singlephoton,qin2015complete_DAVIS22,polycarpou2012adaptive_DAVIS21}.
As the most fundamental measurement techniques in quantum optics,
both homodyne detection  \cite{qin2015complete_DAVIS22,polycarpou2012adaptive_DAVIS21,PhysRevLett.124.213603_Nanshijie} and photon counting  \cite{wasilewski2007spectral_DAVIS20,Thiel:19_FrqRslvHOM_singlephoton,qin2015complete_DAVIS22,polycarpou2012adaptive_DAVIS21} are utilized in these methods. 

The Hong-Ou-Mandel (HOM) interference can be used to analyze temporal-spectral properties of single photons.
HOM interference is the bunching effect when identical photons incident a beam splitter (BS) from different directions and is depicted by the reduction of coincidence rate  \cite{PhysRevLett.59.2044}.
The visibility of HOM interference is affected by the photon statistic feature and indistinguishability between incident photons  \cite{PhysRevLett.118.153603_Disting}, including their polarization, frequency, intensity, etc. 
The change of coincidence rate with respect to relative delay between incident photons is called the HOM interference pattern or HOM dip. 
As the dip is a curve in the time domain, its profile is affected by the temporal mode matching of the incident photons. In the specific case of HOM interference between two independent photons, the profile of the interference pattern is determined by the mode matching degree \cite{PhysRevResearch.4.023125_LI_OU_FOUR_ORDER}, a parameter defined as the square magnitude of the cross-correlation function between their temporal modes. 
Many studies have investigated the impact of mode mismatch on the HOM dip, particularly in scenarios where dispersion is introduced into wave packets in different states, such as the thermal state, single-photon state, and coherent state  \cite{Xiaoxin2015Hong,ma2011effect,fan2021effect}.
However, the extraction of mode information is currently limited to specific and simplified conditions, such as measuring the linear chirp  \cite{Y1993Measurement} and pulse width  \cite{ fan2021effect,2021Propagation} of the incident wave packets in Gaussian modes. Further research is to develop methods for obtaining a complete profile of the temporal mode, transcending the extraction of coefficients associated with specific functions.

In this letter, we present a method to measure the phase spectrum difference (PSD) between two independent wave packets.
We found and utilized the phase retrieval problem \cite{J1982Phase}  inherent in HOM interference, that is, the attempt to reconstruct the PSD from the HOM dip and intensity spectra of the incident wave packets. 
The algorithms we adapted are the Gerchberg-Saxtion (G-S)  algorithm \cite{Gerchberg1972APA} and the generalized projection (GP) algorithm \cite{Yudilevich1987RestorationOS_GPimage}, two of the most basic phase retrieval algorithms that have been used in the field of image recovery \cite{Gerchberg1972APA,Yudilevich1987RestorationOS_GPimage} and frequency-resolved optical gating \cite{Trebino1993UsingPR_Trebino,1994Frequency}, and special modification was employed to suppress the impact of the system noise from photon counting.
The confirmatory experiment was implemented using weak coherent wave packets, where the PSD is introduced by a programmable filter (PF), comparing the reconstructed PSD with the preset PSD, we could gauge the feasibility of our method.
Furthermore, this method allows for the full characterization of an unknown single-photon wave packet by utilizing a fully known wave packet and the single-photon intensity spectrum measurement\cite{Smith2009PhotonPG_MONO_jsi,Davis2016PulsedSS_CHIRPED_JSI}. As the other methods using photon counting systems and interferometry  \cite{wasilewski2007spectral_DAVIS20,Thiel:19_FrqRslvHOM_singlephoton,qin2015complete_DAVIS22,polycarpou2012adaptive_DAVIS21} require the measurement of two-dimensional data, our phase retrieval method, which only requires the collection of one-dimensional data, could speed up the data collection process. 

\section{Theory}


The expression of a single-photon wavepacket can be written as

\begin{equation}
    {\left| 1 \right>_{\psi}} =  \int {\text{d}\omega} \psi{\left( \omega \right)}  \hat{a}^{\dagger}{\left( \omega \right)}  {\left|  \text{vac} \right>},
\end{equation}
where $\psi (\omega) $ is a normalized function representing a specific temporal-spectral mode, with the Fourier transform relation: 
\begin{equation}
    \tilde{\psi}\left( t \right)
    =\frac{1}{\sqrt{2\pi}}
    \int{  \text{d}\omega 
            \psi\left(\omega\right)
            \text{e}^{-\text{i}\omega t}}.
\end{equation}
$\tilde{\psi}\left( t \right)$ is usually seen as the wave function of a single-photon wave packet in the time domain.

\begin{figure} 
    \centering
    \includegraphics[width=8cm]{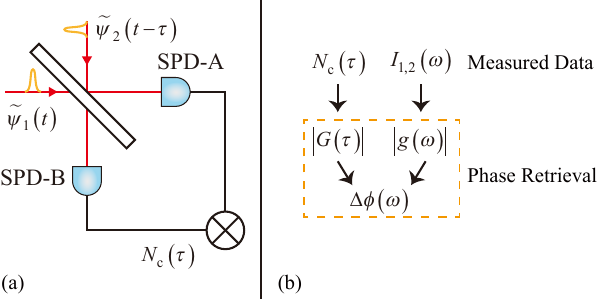} 
    \caption{ (a) Schematic of HOM interference between indepedent photons. Two photons with different temporal mode $\tilde{\psi_1}\left( t \right)$, $\tilde{\psi_2}\left( t-\tau \right)$ incident the BS, from different directions with a relative delay $\tau$, and two SPDs are placed at both of the out port of the BS respectively. Photon detection events will be recorded, allowing the calculation of the normalized coincidence rate $N_\text{c}\left(\tau\right)$. 
    BS, beam splitter; SPD, single photon detector (b) Schematic of the relationship between  the intensity spectra $I_{1,2}\left(\omega\right)$, normalized coincidence rate $N_\text{c}\left(\tau\right)$, the magnitude of cross-correlation function $G\left(\tau\right)$ and  cross-spectral density $g\left(\omega\right)$ and the uncharacterized  PSD $\Delta\phi\left(\omega\right)$. 
    }
    \label{fig:1}
\end{figure}
The HOM interference between independent single-photon wave packets is shown in Fig. \ref{fig:1}(a).
Assuming that the mode in other dimensions, such as the polarizing and transversal spatial modes of the two incident beams, matches perfectly, the possibility of detecting both photons is
\begin{equation}
    {P}_\text{AB} {\left(\tau\right)}= \eta_\text{A}\eta_\text{B}
    \left(
    1-V\left(\tau\right)
    \right),
\end{equation}
where $\eta$ is the efficiency of photon detection
and $V\left(\tau\right)$ is the mode matching degree, defined as
\begin{equation}
    V {\left({\tau}\right)} 
    =  {\left| G\left({\tau}\right) \right|}^2 = {\left| 
    \int \text{d}t \tilde{\psi}^{*}_{1}\left(t\right)
    \tilde{\psi}^{}_{2}\left(t-\tau\right)
    \right|}^2.
\end{equation}
$G(\tau)$ represents the cross-correlation function between the two incident wave packets in the time domain, and $V\left(\tau\right)$ is the square modulus of it.

In this research, the normalized coincidence rate ${N_\text{c}}{\left(\tau\right)}$ is used to depict the interference pattern to better reflect the relation between the interference pattern and temporal modes matching of incident photons. This relation is shown as follows:
\begin{equation}
    {N_\text{c}}{\left(\tau\right)}=\frac{{P}_\text{AB} {\left(\tau\right)}}{{P}_\text{AB} {\left(\infty\right)}} = 1- V{\left(\tau\right)}.
    \label{eq:ncdefine}
\end{equation}
Taking the scenario of coherent wave packets (which are analogous to pulses from a mode-locked laser) into consideration, due to their photon statistical characteristics, it is possible that both SPDs detected a photon from the same incident port,
decreasing the global visibility of HOM interference. The factor of $V(\tau)$ in Eq. (\ref{eq:ncdefine}) should be modified to $1/2$ when both the incident wave packets are in the coherent state with the same magnitude and $2/(\left|A\right|^2+2)$ when interference happens between a single-photon wave packet and a coherent one, with $|A|$ as the relative magnitude. The detailed derivation for all these conditions mentioned above is provided in Appendix \ref{apdxA}, and we can find that $V(\tau)$, as the square modulus of the cross-correlation function  $G(\tau)$, can be measured from photon counting in HOM interference.


According to the generalized Wiener-Khinchin theorem, 
 $G(\tau)$ can be rephrased in the integration in the frequency domain as
\begin{align}
    G\left({\tau}\right) 
    & = \int{\text{d}\omega \psi _{1}^{*}\left( \omega  \right)\psi _{2}^{{}}\left( \omega  \right){{e}^{i\omega \tau }}} \notag \\
    & = \int{  \text{d}\omega g\left({\omega}\right) {e}^{i\omega \tau }},
\end{align}
where  $g\left(\omega\right)$ is the cross-spectral density. 
Rewriting $\psi(\omega)$ and  $g(\omega)$ in terms of magnitude and argument, we have:
\begin{equation}\psi(\omega)= 
    \left| \psi\left({\omega}\right) \right|
    {e}^{i\phi \left( \omega  \right)},
\end{equation} 
\begin{equation}
    g\left({\omega}\right) = 
    \left| g\left({\omega}\right) \right|
    {e}^{i\Delta \phi \left( \omega  \right)}.
\end{equation}
$\phi \left( \omega  \right)$ is the phase spectrum of $\psi \left( \omega  \right)$ and $\Delta \phi \left( \omega  \right)$ is the phase spectrum difference(PSD), defined as $\Delta \phi \left( \omega  \right) =\phi_2 \left( \omega  \right)-\phi_1 \left( \omega  \right)$.
In this paper, we regard $\left| g\left({\omega}\right) \right|$ as a measurable function as it can be characterized by measuring the intensity spectrum of the two incident wave-packets, as $\left| g\left({\omega}\right) \right| = \sqrt{I_1\left(\omega\right)I_2\left(\omega\right)}$.

According to the previous discussion,  the magnitudes of both functions in the Fourier transform pair, $g\left({\omega}\right)$ and $G\left({\tau}\right)$, are directly measurable, while the phase terms of this pair cannot be measured. The challenge of reconstructing the phase from 
the intensities of the two functions that forms a Fourier transform pair is called the phase retrieval problem  \cite{J1982Phase}.
Following this concept of phase retrieval, the PSD $\Delta \phi \left( \omega  \right)$ is possible to be reconstructed  from the spectrum $I_{1,2}(\omega)$ and the interference pattern $N_\text{c}(\tau)$, as is shown in Fig. \ref{fig:1}(b).

We used a composite algorithm to finish the phase recovery process, including the G-S algorithm\cite{Gerchberg1972APA} and the GP algorithm\cite{Yudilevich1987RestorationOS_GPimage}. We implemented a little adaption on them to reduce the impact of noise in the last few iterations.
\begin{figure}[b]    \centering\includegraphics[width=5cm]{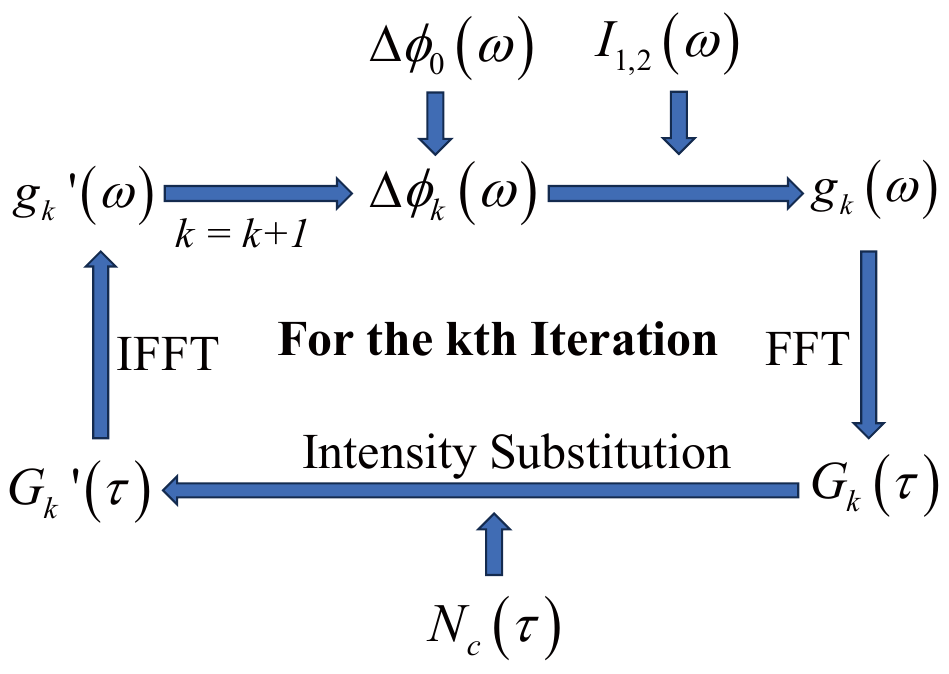}
    \caption{Principle of the G-S algorithm. 
    The functions with a $k$ subscript represent the current guess of cross correlation function $G\left(\tau\right)$ and cross-spectral density $g\left(\omega\right)$. 
    The iteration starts with a initial guess of phase spectrum difference $\Delta\phi\left(\omega\right)$. Intensity spectrua $I_{1,2}\left(\omega\right)$ 
    and normalized coincidence rate
    $N_\text{c}\left(\tau\right)$
    are used to correct the guess functions in their domain respectively
    .
    FFT, fast Fourier transform; IFFT, inverse fast Fourier transform}
    \label{fig:2}
\end{figure}
These basic algorithms follow a similar process, which requires iterative Fourier transform between the two domains. Iteration procedure of the G-S algorithm is shown in Fig. \ref{fig:2} as example.
Specific iteration procedures with formulas and simulation results can be found in Appendix \ref{apdxB}.
\section{experiment}
Experiments were implemented to demonstrate the availability of our method.
We chose the interference between coherent  wave packets for the confirmation experiment for the following reasons: first, the source, readily available from a mode-locked laser, can be directed through attenuators (ATT) to achieve the single-photon level required for photon counting detection; second, conversely, the spectrum of the coherent source can be directly measured using a spectrometer when the ATTs are removed from the optical path, offering a more convenient alternative to direct measurement at the single-photon level.

\begin{figure}[htbp]    \centering\includegraphics[width=7cm]{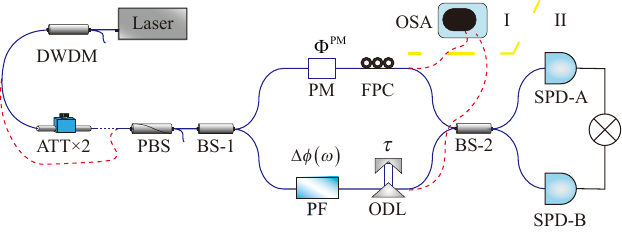}
    \caption{Experimental setup. DWDM, dense wavelength division multiplexer; ATT, attenuator; PBS, polarization beam splitter; BS, beam splitter; PZT, piezoelectric ceramic transducer; FPC, fiber polarization controller; OSA, optical spectrum analyzer; PF, programmable filter; ODL, optical delay line; SPD, single photon detector}
    \label{fig:3}
\end{figure}

Fig. \ref{fig:3} illustrates our experimental setup, where a mode-locked fiber laser generates coherent  wave packets, filtered by a dense wavelength division multiplexer to produce a 1.2 nm-wide spectrum. Passing through two ATTs, the intensity of the  wave packets is attenuated for photon-counting detection, ensuring a direct proportionality between the counting rate and intensity. A polarization beam splitter then polarizes the  wave packets.
On the signal arm, a Finisar 4000A WaveShaper is functioning as the PF to apply an arbitrary phase function $\Delta\phi_A(\omega)$,  which corresponds to the PSD. On the reference arm, an electronic translation stage with a right-angle prism serves as the optical delay line. This setup allows for precise and programmable control over the relative delay $\tau$. Additionally, a Piezoelectric Transducer (PZT) is positioned as a phase modulator (PM). 
The PZT is driven by a periodic triangular voltage signal, resulting in a phase modulation with an amplitude corresponding to a $6\pi$ phase shift, to ensure that  $\left\langle \text{exp}\left({i\Phi^{\text{PM}}}\right)\right\rangle_{\text{pulses}}=0$ over a long period of counting time. Consequently, the  wave packets meeting at BS-2 are effectively phase-independent.
According to the experiment set up, the interference pattern is
\begin{equation}
            N_\text{c}\left( \tau  \right)=1-\frac{1}{2}{{\left| \int{\text{d}\omega \sqrt{I_1\left( \omega  \right)I_2\left( \omega  \right)}{{e}^{i\Delta \phi \left( \omega  \right)}}{{e}^{i\omega \tau }}} \right|}^{2}}.
\end{equation}
The PSD $\Delta \phi \left( \omega  \right)$ is considered equal to $\Delta \phi_\text{A} \left( \omega  \right)$ under the condition that the two wave packets overlap at $\tau = 0$.

In the experiment, we first applied a phase function by the PF, and then measured the spectrum and interference pattern. Our algorithm can produce a reconstructed  PSD $\Delta \phi_k \left( \omega  \right)$ with the measured data. Comparing it with $\Delta \phi_\text{A} \left( \omega  \right)$, we can assess the applicability of this method.
The measurement is divided into two steps:
In Step I, with the ATTs not engaged in the optical path, the spectra of the  wave packets on both arms need to be measured, as the PF may introduce a significant insertion loss spectrum, particularly when the slope of $\Delta\phi_A(\omega)$ is steep, resulting in a difference in the spectra on the two arms. The magnitude of $g(\omega)$ is then characterized by $\left|g\left(\omega\right)\right|=\sqrt{I_1\left(\omega\right)I_2\left(\omega\right)}$.
In Step II, pairs of  wave packets at the single-photon level incident on BS-2, which is connected to two SPDs, where counting rates will be recorded at different relative delay $\tau$ and a normalized coincidence $N_{\text{c}}\left(\tau\right)$ rate will be calculated to derive the magnitude of $G(\tau)$ with $\left|G(\tau)\right| =\sqrt{2 -2 N_{\text{c}}\left(\tau\right)}$.

Enhancing the counting rate and sampling duration can raise the accuracy of measured patterns.
However, if the counting rate is too high, the nonlinear relationship between the counting rate and intensity will be obvious, resulting in a reduced expected value for the single-channel counting rate and an increased expected value for the normalized coincidence rate.
In order to ensure the accuracy and efficiency of the measurement, in our experiment, the counting rate of SPD is about 1.3 MHz, while the repetition rate of the laser source is about 36.88 MHz, and dead time of SPD is about 80 ns on average. 
It takes 30 s to measure the $N_\text{c}$ at each sampling point for the first experiment and 90 s for the second one.

\begin{figure}[htbp]    \centering\includegraphics[width=8cm]{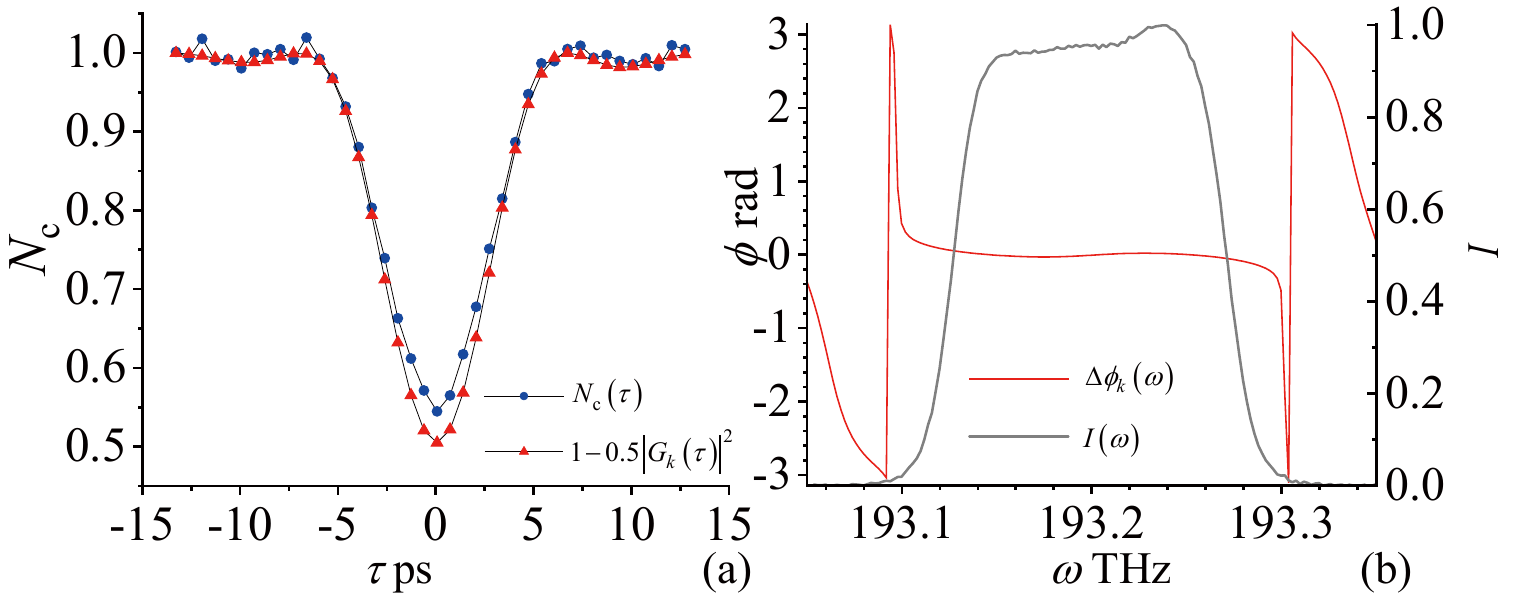}
    \caption{Experiment result when $\Delta {{\phi }_{A}}\left( \omega  \right)=0$. 
    (a) Comparison between $\left|G_{k} \left(\tau\right)\right|^2$ in triangular dotted red line and $N_\text{c}\left( \tau \right)$ in circle dotted blue line.
    (b) Comparison between reconstructed  phase function $\Delta\phi_k(\omega)$ in solid red line and and equivalent spectrum $I \left(\omega \right) $ in gray line.}
    \label{fig:4}
\end{figure}

Fig. 
\ref{fig:4} shows the HOM dip 
and reconstructed  phase function when $\Delta {{\phi }_{A}}\left( \omega  \right)=0$. As shown in Fig. \ref{fig:4}(b), the reconstructed phase function $\Delta\phi_k(\omega)$ is the smooth curve close to zero at frequencies with strong intensity. As the intensity spectrum decreases at frequency outside the central region, the phase function increasingly resembles a random value. This randomness is due to the minimal contribution of the weaker intensity frequency components to the interference pattern, leading to less accurate reconstruction of the phase at corresponding frequency.

Unmatched polarization and intensity differences between the incident  wave packets will reduce the global visibility of the interference pattern, which means that the overall amplitude of the measured function $\left|G(\tau)\right|$ will be smaller. 
However, as our algorithm is an iterative Fourier transform method, the overall amplitude wouldn't affect the $\Delta\phi_k(\omega)$. Reversely, The iteration outcomes can be used to correct the measured HOM dip.
The lowest point of the reconstructed  HOM dip is 0.505, which is close to 0.5, indicating that the different of dispersion between the two arms due to the length difference in fibers and other elements are negligible.
 Furthermore, we manually aligned the lowest point of the HOM dip with  $\tau=0$. This manual calibration serves as a reference for the relative delay, assisting in the detrending of the reconstructed  phase function. This deliberate calibration helps to reduce the need for further manual manipulation when making comparisons, allowing for a more scientific and less subjective comparison between $\Delta {{\phi }_{A}}\left( \omega  \right)$ and $\Delta\phi_k(\omega)$.

\begin{figure}[htbp]    \centering\includegraphics[width=8cm]{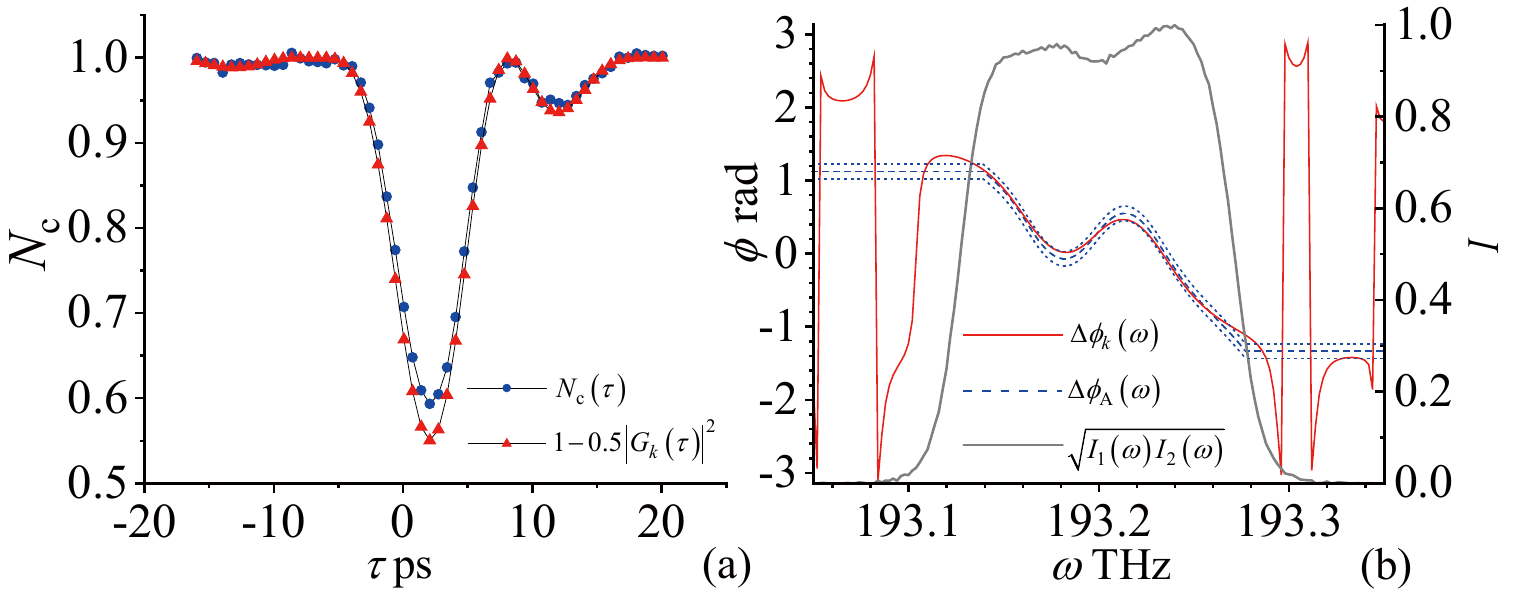}
    \caption{Experiment result when applied an "inverted N-shaped" function on PF.  
    (a) Comparison between $\left|G_{k} \left(\tau\right)\right|^2$ in triangular dotted red line and $N_\text{c}\left( \tau \right)$ in circle dotted blue line.
    (b) Comparison between reconstructed  phase function $\Delta\phi_k(\omega)$ in solid red line and applied phase function $\Delta\phi_A(\omega)$ in dash blue line, with the equivalent spectrum $\sqrt{I_1 \left(\omega \right)I_2 \left(\omega \right)}$ in gray line.}
    \label{fig:5}
\end{figure}

Fig. \ref{fig:5} shows the HOM dip and reconstructed  phase function when introducing an "inverted N-shaped" function $\Delta {{\phi }_{A}}\left( \omega  \right)$.  
Prior to the experiment, we simulated the effects of counting noise by incorporating the binomial distribution characteristic of single-channel counting. Our simulations revealed that the majority of the reconstructed  $\Delta\phi_k(\omega)$ fall within the interval  
$\Delta\phi_A(\omega)\pm0.1$ $\text{rad}$ between 193.13 $\sim$ 193.25 THz. As the reconstructed  PSD in Fig. \ref{fig:5}(b) is in this confidence interval, the noise from the counting distribution dominates the error, proving that our method can successfully reconstruct the PSD $\Delta\phi(\omega)$ between the two incident wave packets.

In both simulation and experiment, we found that the specific phase retrieval problem we processed presents a double solution problem: the reconstructed  $\Delta\phi_k(\omega)$ may converge to $-\Delta\phi_A(\omega_0-\omega)$, which is centrally symmetric to $\Delta\phi_A(\omega)$, this phenomenon is particularly evident when the spectrum is symmetric or the noise is high. However, the mode matching degree corresponding to the incorrect convergence results $\left|G_{k} \left(\tau\right)\right|^2$ will show a greater difference compared to the correct convergence results when compared to the actual mode matching degree $V\left( \tau \right)$. Therefore, we run the algorithm with varied initial values and compare the results to address the double solution issue.

The scenario in our last experiment can be likened to 
measuring the phase spectrum of an unknown single-photon wave packet from its intensity spectrum and HOM dip with a known reference wave packet in the coherent state, as the interference pattern is still determined by $V\left(\tau\right)$. Ideally, the intensity spectrum of the reference wave packet should be marginally wider than that of the unknown one to fully reconstruct the phase spectrum.

\section{summary}
In summary, we present a method for reconstructing the PSD between two wave packets at the single photon level from their HOM dip and intensity spectra, 
by an adapted phase retrieval algorithm. 
Our composite algorithm integrates the G-S and GP algorithms, and is modified to reduce the impact of system noise inherent in the photon counting system.
To assess the effectiveness of our method, we performed experiments using coherent  wave packets. 
The experiment results indicate that the reconstructed PSD 
 $\Delta\phi_k(\omega)$ approximated the applied PSD $\Delta \phi_A(\omega)$ introduced by the PF, with a discrepancy within ±0.1 rad.
Our method is also capable of fully characterizing the single-photon wave packet with a known wave packet,  requiring only one-dimensional data, that is the intensity spectra of the wave packets and the HOM dip. 
However, it should be noted that this method does not facilitate the determination of a density matrix for undefined states, making it only suitable for the characterization of pure states rather than mixed states.

\begin{acknowledgments}
We wish to acknowledge Professor Zhe-Yu Jeff Ou for constructive discussions.
\end{acknowledgments}

\appendix
\section{Equation Derivation about HOM interference}\label{apdxA}

Here, we provide detailed derivations of the HOM dip expression used in the main text, for incident wavepackets in different states. As mentioned in the main text, 
the single-photon wave packet is
\begin{equation}
    \left| 1 \right\rangle =\int{\text{d}\omega \psi \left( \omega  \right){\text{e}^{\text{i}\omega {{t}_{0}}}}\hat{a}_{{}}^{\dagger }\left( \omega  \right)\left| 0 \right\rangle },
    \label{eq:A1}
\end{equation}
where $\hat{a}_{{}}^{\dagger }\left( \omega  \right)$ represents the creation operator at $\omega$
and the expression of the electromagnetic operator is
\begin{equation}
\hat{E}^{(+)}\left(t\right)=\frac{1}{\sqrt{2\pi }}\int{\text{d}\omega \hat{a}\left( \omega  \right){{\text{e}}^{-\text{i}\omega t}}}.
\end{equation}
Applying it on the single-photon wave packet, we have
\begin{equation}
  \hat{E}^{(+)}\left( t \right)\left| 1 \right\rangle =\tilde{\psi }\left( t-{{t}_{0}} \right)\left| 0 \right\rangle  .
\end{equation}
The coherent state wave packet is a coherent superposition of single-mode coherent states as
\begin{equation}
    \left| \alpha \right\rangle =\int{\text{d}\omega \psi \left( \omega  \right){\text{e}^{\text{i}\omega {{t}_{0}}}}\hat{a}^{\dagger }\left( \omega  \right)\left| \alpha\left( \omega \right)\right\rangle }.
\end{equation}
Applying an Electromagnetic operator on it, we get
\begin{equation}
\hat{E}_{{}}^{(+)}\left( {{t}_{}} \right) \left| \alpha  \right\rangle =A\tilde{\psi }\left( t-{{t}_{0}} \right)\left| \alpha  \right\rangle .
\label{eq:A5}
\end{equation}
According to Glauber's theory of detection, the possibility of detecting a photon when a wave packet arrive the detector is
\begin{equation}
          {{P}_{}}={{\eta }_{}}\int{\text{d}t\left\langle  \psi  \right|}\hat{E}_{}^{(-)}\left( t \right)\hat{E}_{}^{(+)}\left( t \right)\left| \psi  \right\rangle  ,
\end{equation}
where $\eta$ is the efficient of the detector. And the possibility of two detectors detect a photon is
\begin{align}
  {P}_\text{AB}= \eta_\text{A}\eta_\text{B}
  \int{\text{d}t_\text{A}t_\text{B}}&
  \left\langle 
  \alpha_1,\alpha_2
  \right|
  \hat{E}_\text{A}^{(-)}(t_\text{A})
  \hat{E}_\text{B}^{(-)}(t_\text{B})
  \notag\\&
  \hat{E}_\text{B}^{(+)}(t_\text{A})
  \hat{E}_\text{A}^{(+)}(t_\text{B})
  \left|
  \alpha_1,\alpha_2
  \right\rangle.
  \label{eq:A7}
\end{align}
To derive the relationship between the HOM dip and the temporal mode of the incident wave packets, we utilize the equations of a beam splitter in the Heisenberg picture. Here, we apply an idealized model of a beam splitter, represented by the following equations:
\begin{equation}
    \hat{E}_\text{A,B}^{(+)}(t)=\frac{\sqrt{2}}{2}\left( \hat{E}_{1}^{(+)}\left( t \right)\pm\hat{E}_{2}^{(+)}\left( t -\tau\right) \right).
\end{equation}

The demonstration of the HOM dip between two single-photon wave packets requires the derivation of ${P}_\text{AB}\left(\tau\right)$.
When one or both of the incident beams are coherent wave packets, the ensemble average $\left\langle{P}_\text{AB}\left(\tau\right)\right\rangle$ is required.
This derivation involves the use of the expressions $\left\langle A \right\rangle^n = 0$ and $\left\langle A_1A_2 \right\rangle^n = 0$. The later one is applicable when the phase of $A_1$ and $A_2$ are independent of each other.
Applying Eqs. (\ref{eq:A1} $\sim$ \ref{eq:A5}) to Eq. (\ref{eq:A7}),
the HOM dip for the aforementioned three conditions can be derived as follows:
\begin{widetext}
\begin{align}
{P}_\text{AB}\left(\tau\right) & = 
\frac{1}{4}\eta_\text{A}\eta_\text{B}
  \int{\text{d}t_\text{A}t_\text{B}}
    \left\langle 1 \right|_1\left\langle 1 \right|_2
    \left( \hat{E}_{1}^{(-)}\left( t_\text{A} \right)+\hat{E}_{2}^{(-)}\left( t_\text{A} -\tau\right) \right)
    \left( \hat{E}_{1}^{(-)}\left( t_\text{B} \right)-\hat{E}_{2}^{(-)}\left( t_\text{B} -\tau\right) \right)
  \notag \\& \phantom{=}
    \left( \hat{E}_{1}^{(+)}\left( t_\text{B} \right)-\hat{E}_{2}^{(+)}\left( t_\text{B} -\tau\right) \right)
  \left( \hat{E}_{1}^{(+)}\left( t_\text{A} \right)+\hat{E}_{2}^{(+)}\left( t_\text{A} -\tau\right) \right)
 \left| 1 \right\rangle_2
    \left| 1 \right\rangle_1  \notag \\
  &=
  \frac{1}{4}\eta_\text{A}\eta_\text{B} 
  \int{\text{d}t_\text{A}t_\text{B}}
    \left\langle 1 \right|_1
    \left\langle 1 \right|_2
        \hat{E}_{1}^{(-)} \left( t_\text{A} \right) 
        \hat{E}_{2}^{(-)} \left( t_\text{B}-\tau \right) 
        \hat{E}_{2}^{(+)} \left( t_\text{B}-\tau \right) 
        \hat{E}_{1}^{(+)} \left( t_\text{A} \right)
\notag  \\ & \phantom{=} +
        \hat{E}_{2}^{(-)} \left( t_\text{A}-\tau \right) 
        \hat{E}_{1}^{(-)} \left( t_\text{B} \right) 
        \hat{E}_{1}^{(+)} \left( t_\text{B} \right)
        \hat{E}_{2}^{(+)} \left( t_\text{A}-\tau \right) 
\notag  \\ & \phantom{=} -
        \hat{E}_{1}^{(-)} \left( t_\text{A} \right) 
        \hat{E}_{2}^{(-)} \left( t_\text{B}-\tau \right) 
        \hat{E}_{1}^{(+)} \left( t_\text{B} \right)
        \hat{E}_{2}^{(+)} \left( t_\text{A}-\tau \right) 
\notag \\ & \phantom{=} -
        \hat{E}_{2}^{(-)} \left( t_\text{A}-\tau \right) 
        \hat{E}_{1}^{(-)} \left( t_\text{B} \right) 
        \hat{E}_{2}^{(+)} \left( t_\text{B}-\tau \right)
        \hat{E}_{1}^{(+)} \left( t_\text{A} \right) 
    \left| 1 \right\rangle_2
    \left| 1 \right\rangle_1
\notag \\ & = 
\frac{1}{4}\eta_\text{A}\eta_\text{B} 
  \int{\text{d}t_\text{A}t_\text{B}}[
  \tilde{\psi }_1^{*}\left(t_\text{A}\right)
  \tilde{\psi }_2^{*}\left(t_\text{B}-\tau\right)
  \tilde{\psi }_2^{}\left(t_\text{B}-\tau\right)
  \tilde{\psi }_1^{}\left(t_\text{A}\right)
  +
  \tilde{\psi }_2^{*}\left(t_\text{A}-\tau\right)
  \tilde{\psi }_1^{*}\left(t_\text{B}\right)
  \tilde{\psi }_1^{}\left(t_\text{B}\right)
  \tilde{\psi }_2^{}\left(t_\text{A}-\tau\right)
  \notag \\ & \phantom{=} -
  \tilde{\psi }_1^{*}\left(t_\text{A}\right)
  \tilde{\psi }_2^{*}\left(t_\text{B}-\tau\right)
  \tilde{\psi }_1^{}\left(t_\text{B}\right)
  \tilde{\psi }_2^{}\left(t_\text{A}-\tau\right)
  -
  \tilde{\psi }_2^{*}\left(t_\text{A}-\tau\right)
  \tilde{\psi }_1^{*}\left(t_\text{B}\right)
  \tilde{\psi }_2^{}\left(t_\text{B}-\tau\right)
  \tilde{\psi }_1^{}\left(t_\text{A}\right)
  ] \notag \\
  & = 
  \frac{1}{2}\eta_\text{A}\eta_\text{B}  
  \left(
  1-\int{\text{d}t}
  \left|
    \tilde{\psi }_1^{*}\left(t\right)
    \tilde{\psi }_2^{}\left(t-\tau\right)
  \right|^{2}
  \right),
 \label{eq: derivation dip 1}
\end{align}

\begin{align}
\left\langle{P}_\text{AB}\left(\tau\right) \right\rangle
  &=
  \frac{1}{4}\eta_\text{A}\eta_\text{B} 
  \int{\text{d}t_\text{A}t_\text{B}}
    \left\langle \alpha \right|_1
    \left\langle 1 \right|_2
        \hat{E}_{1}^{(-)} \left( t_\text{A} \right) 
        \hat{E}_{2}^{(-)} \left( t_\text{B}-\tau \right) 
        \hat{E}_{2}^{(+)} \left( t_\text{B}-\tau \right) 
        \hat{E}_{1}^{(+)} \left( t_\text{A} \right)
\notag  \\ & \phantom{=} +
        \hat{E}_{2}^{(-)} \left( t_\text{A}-\tau \right) 
        \hat{E}_{1}^{(-)} \left( t_\text{B} \right) 
        \hat{E}_{1}^{(+)} \left( t_\text{B} \right)
        \hat{E}_{2}^{(+)} \left( t_\text{A}-\tau \right) 
\notag \\ & \phantom{=} -
        \hat{E}_{1}^{(-)} \left( t_\text{A} \right) 
        \hat{E}_{2}^{(-)} \left( t_\text{B}-\tau \right) 
        \hat{E}_{1}^{(+)} \left( t_\text{B} \right)
        \hat{E}_{2}^{(+)} \left( t_\text{A}-\tau \right) 
\notag \\ & \phantom{=} -
        \hat{E}_{2}^{(-)} \left( t_\text{A}-\tau \right) 
        \hat{E}_{1}^{(-)} \left( t_\text{B} \right) 
        \hat{E}_{2}^{(+)} \left( t_\text{B}-\tau \right)
        \hat{E}_{1}^{(+)} \left( t_\text{A} \right)
\notag \\ & \phantom{=} +
       \hat{E}_{1}^{(-)} \left( t_\text{A} \right) 
        \hat{E}_{1}^{(-)} \left( t_\text{B} \right) 
        \hat{E}_{1}^{(+)} \left( t_\text{B} \right)
        \hat{E}_{1}^{(+)} \left( t_\text{A} \right)
    \left| 1 \right\rangle_2
    \left| \alpha \right\rangle_1
\notag \\ & = 
  \frac{1}{4}\eta_\text{A}\eta_\text{B}  
  \left(
  2\left|A\right|^2 -  2\left|A\right|^2\int{\text{d}t}
  \left|
    \tilde{\psi }_1^{*}\left(t\right)
    \tilde{\psi }_2^{}\left(t-\tau\right)
  \right|^{2}
  +\left|A\right|^4
  \right),
 \label{eq: derivation dip 2}
\end{align}
\begin{align}
\left\langle{P}_\text{AB}\left(\tau\right) \right\rangle
  &=
  \frac{1}{4}\eta_\text{A}\eta_\text{B} 
  \int{\text{d}t_\text{A}t_\text{B}}
    \left\langle \alpha \right|_1
    \left\langle \alpha \right|_2
        \hat{E}_{1}^{(-)} \left( t_\text{A} \right) 
        \hat{E}_{2}^{(-)} \left( t_\text{B}-\tau \right) 
        \hat{E}_{2}^{(+)} \left( t_\text{B}-\tau \right) 
        \hat{E}_{1}^{(+)} \left( t_\text{A} \right)
\notag  \\ & \phantom{=} +
        \hat{E}_{2}^{(-)} \left( t_\text{A}-\tau \right) 
        \hat{E}_{1}^{(-)} \left( t_\text{B} \right) 
        \hat{E}_{1}^{(+)} \left( t_\text{B} \right)
        \hat{E}_{2}^{(+)} \left( t_\text{A}-\tau \right) 
\notag \\ & \phantom{=} -
        \hat{E}_{1}^{(-)} \left( t_\text{A} \right) 
        \hat{E}_{2}^{(-)} \left( t_\text{B}-\tau \right) 
        \hat{E}_{1}^{(+)} \left( t_\text{B} \right)
        \hat{E}_{2}^{(+)} \left( t_\text{A}-\tau \right) 
\notag \\ & \phantom{=} -
        \hat{E}_{2}^{(-)} \left( t_\text{A}-\tau \right) 
        \hat{E}_{1}^{(-)} \left( t_\text{B} \right) 
        \hat{E}_{2}^{(+)} \left( t_\text{B}-\tau \right)
        \hat{E}_{1}^{(+)} \left( t_\text{A} \right)
\notag \\ & \phantom{=} +
       \hat{E}_{1}^{(-)} \left( t_\text{A} \right) 
        \hat{E}_{1}^{(-)} \left( t_\text{B} \right) 
        \hat{E}_{1}^{(+)} \left( t_\text{B} \right)
        \hat{E}_{1}^{(+)} \left( t_\text{A} \right)
\notag \\ & \phantom{=} +
    \hat{E}_{2}^{(-)} \left( t_\text{A}-\tau \right) 
        \hat{E}_{2}^{(-)} \left( t_\text{B}-\tau \right) 
        \hat{E}_{2}^{(+)} \left( t_\text{B}-\tau \right)
        \hat{E}_{2}^{(+)} \left( t_\text{A}-\tau \right)
    \left| \alpha \right\rangle_2
    \left| \alpha \right\rangle_1
\notag \\ & = 
  \frac{1}{4}\eta_\text{A}\eta_\text{B}  
  \left(
  2\left|A_1\right|^2\left|A_2\right|^2 -  2\left|A_1\right|^2\left|A_2\right|^2
  \int{\text{d}t}
  \left|
    \tilde{\psi }_1^{*}\left(t\right)
    \tilde{\psi }_2^{}\left(t-\tau\right)
  \right|^{2}
  +\left|A_1\right|^4
  +\left|A_2\right|^4
  \right).
 \label{eq: derivation dip 3}
\end{align}
\end{widetext}
The corresponding normalized coincidence rate is
\begin{equation}
    {N_\text{c}}{\left(\tau\right)} = 1- V{\left(\tau\right)},
\end{equation}
\begin{equation}
    N_\text{c}\left(\tau\right) = 1 - \frac{2}{(\left|A\right|^2+2)}V\left(\tau\right),
\end{equation}
\begin{equation}
    {N_\text{c}}{\left(\tau\right)} = 1- \frac{2\left|A_1\right|^2\left|A_2\right|^2}{2\left|A_1\right|^2\left|A_2\right|^2
    +\left|A_1\right|^4
  +\left|A_2\right|^4}V{\left(\tau\right)}.
\end{equation}

\section{The iteration procedures of the algorithms and the simulation result}\label{apdxB}

As is shown in Fig. \ref{fig:2}, the G-S algorithm iterates in the following steps,
\begin{equation}
    \left\{ 
    \begin{array}{lr}
{{G}_{k}}\left( \tau  \right)=\int{{{g}_{k}}\left( \omega  \right)\exp \left( -\text{i}\omega \tau  \right)\text{d}\omega }
        &\\
{{G}_{k}}'\left( \tau  \right)={\left| {{G}_{}}\left( \tau  \right) \right|}{{{G}_{k}}\left( \tau  \right)}/{\left| {{G}_{k}}\left( \tau  \right) \right|}
        &\\
{{g}_{k}}'\left( \omega  \right)={1}/{2\pi }\int{{{G}_{k}}'\left( \tau  \right)\exp \left( \text{i}\omega \tau  \right)\text{d}\tau }
         &\\
{{g}_{k+1}}\left( \omega  \right)=
{\left| g\left( \omega  \right) \right|}
{{{g}_{k}}'\left( \omega  \right)}/{\left| {{g}_{k}}'\left( \omega  \right) \right|}
         &
    \end{array}
    \right.
    \label{eq:B1}
\end{equation}
which can be summarized as two Fourier transformations and two substitution steps. In practice, the transformation of discrete data was performed by the fast Fourier transform algorithm.

The GP algorithm follows a similar procedure to the G-S algorithm. However, the step generating ${{g}_{k+1}}\left( \omega_i \right)$ is replaced by a gradient descent step. ${{g}_{k+1}}\left( \omega_i \right)$ is defined as
\begin{equation}
    {{g}_{k+1}}\left( \omega_i \right)
    =
   \left| g\left( \omega_i  \right) \right|{\text{{e}}^{\text{i}\Delta \phi \left( {{\omega }_{i}} \right)}},
\end{equation}
 where $\phi \left( {{\omega }_{i}} \right)$ are undetermined coefficients. The corresponding distance function is given by:
\begin{equation}
    Z=\sum\limits_{i}{{{\left( 
    \left| g\left( \omega_i  \right) \right| 
    {\text{{e}}^{\text{i}\Delta \phi \left( {{\omega }_{i}} \right)}}-{{g}_{k}}^{\prime }\left( 
{{\omega }_{i}} \right) \right)}^{2}}}.
\end{equation}
The calculation of the corresponding partial derivatives is presented in Eq. (\ref{eq:B4}).

In the HOM interference, the counting noise is positively correlated to the counting rate itself, which means that in the shallow part of the HOM dip, where $\left|G\left(\tau\right)\right|$ is small, the signal-to-noise ratio does not decrease accordingly. To reduce the impact of counting noise, in the final iterations of our algorithm, we used an adapted G-S algorithm, where
the second line in Eqs. (\ref{eq:B1}) is replaced by Eq. (\ref{eq:B5}) for the points where $\left| {G}\left( \tau_i  \right) \right|$ is smaller than a certain value.
\begin{widetext}
    \begin{equation}
    {{\left. \frac{\partial Z}{\partial \Delta \phi \left( {{\omega }_{i}} \right)} \right|}_{\Delta \phi \left( {{\omega }_{i}} \right)=\Delta {{\phi }_{k}}\left( {{\omega }_{i}} \right)}}=2\left| {{g}_{k}}^{\prime }\left( {{\omega }_{i}} \right) \right|I\left( {{\omega }_{i}} \right)\sin \left( \Delta \phi \left( {{\omega }_{i}} \right)-{{\Delta \phi }_{k}}^{\prime }\left( {{\omega }_{i}} \right) \right).
    \label{eq:B4}
\end{equation}
    \begin{equation}
     {{G}_{k}}^{\prime }\left( \tau  \right)
    =
        \left(
            \frac{1}{2} \left| G\left(\tau\right)\right|^2
            \right.=
            \left. +
            \left( 1-\frac{1}{2} \left| G\left(\tau\right)\right| \right) \left|                     {G_k}\left(\tau\right) \right|
        \right)
        \frac{{{G}_{k}}\left( \tau  \right)}{\left| {{G}_{k}}\left( \tau  \right)\right|}.
        \label{eq:B5}
\end{equation}
\end{widetext}

\section{simulation result}
\begin{figure}
    \centering
    \includegraphics[width = 8cm]{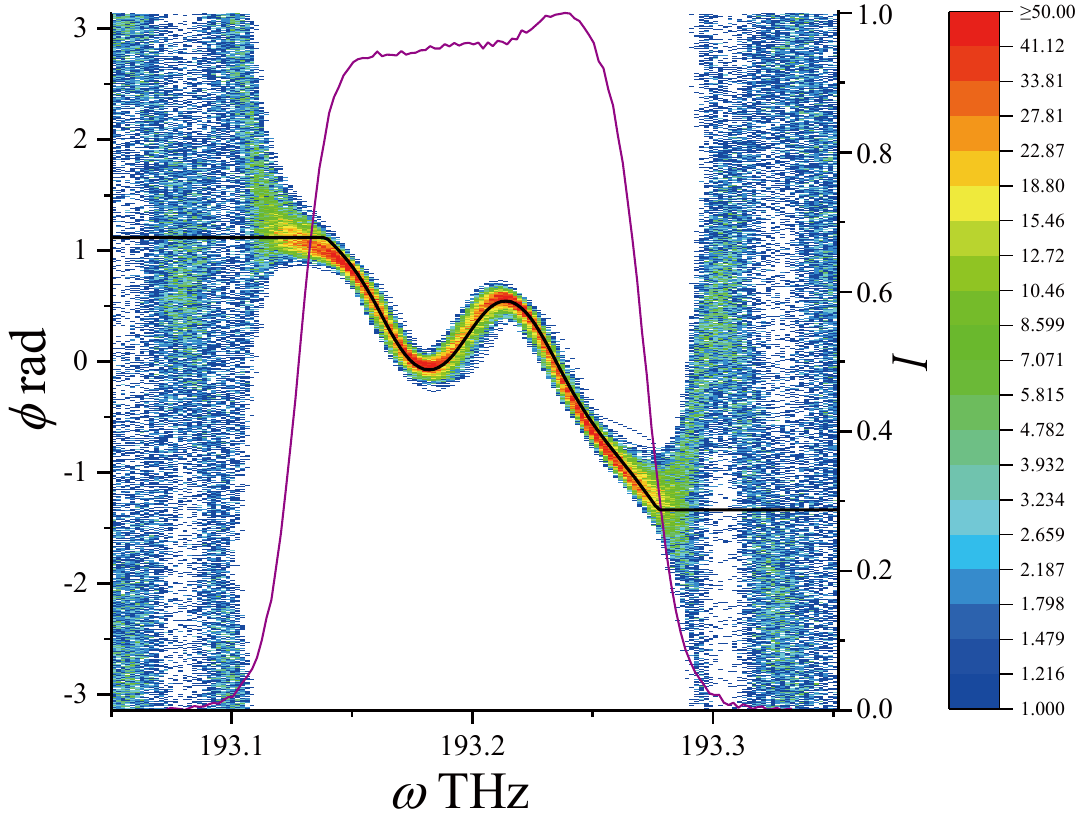}
    \caption{Simulation result. The black solid line is the preset PSD for reference. The heat map shows the distribution of reconstructed phase at different frequencies.}
    \label{noise simulation}
\end{figure}

To assess algorithm convergence amidst noisy interference patterns, we conducted a simulation that takes into account the binomial distribution characteristic of photon counting. We executed 1000 independent simulations with parameters identical to those used in our second experiment. From the noisy interference pattern and the spectra, we successfully reconstructed  the phase functions.  The heatmap corresponding to the simulation results is displayed in Fig. \ref{noise simulation}. These outcomes serve as benchmarks for establishing confidence region.

\bibliography{apssamp}

\begin{thebibliography}{31}%
\makeatletter
\providecommand \@ifxundefined [1]{%
 \@ifx{#1\undefined}
}%
\providecommand \@ifnum [1]{%
 \ifnum #1\expandafter \@firstoftwo
 \else \expandafter \@secondoftwo
 \fi
}%
\providecommand \@ifx [1]{%
 \ifx #1\expandafter \@firstoftwo
 \else \expandafter \@secondoftwo
 \fi
}%
\providecommand \natexlab [1]{#1}%
\providecommand \enquote  [1]{``#1''}%
\providecommand \bibnamefont  [1]{#1}%
\providecommand \bibfnamefont [1]{#1}%
\providecommand \citenamefont [1]{#1}%
\providecommand \href@noop [0]{\@secondoftwo}%
\providecommand \href [0]{\begingroup \@sanitize@url \@href}%
\providecommand \@href[1]{\@@startlink{#1}\@@href}%
\providecommand \@@href[1]{\endgroup#1\@@endlink}%
\providecommand \@sanitize@url [0]{\catcode `\\12\catcode `\$12\catcode `\&12\catcode `\#12\catcode `\^12\catcode `\_12\catcode `\%12\relax}%
\providecommand \@@startlink[1]{}%
\providecommand \@@endlink[0]{}%
\providecommand \url  [0]{\begingroup\@sanitize@url \@url }%
\providecommand \@url [1]{\endgroup\@href {#1}{\urlprefix }}%
\providecommand \urlprefix  [0]{URL }%
\providecommand \Eprint [0]{\href }%
\providecommand \doibase [0]{https://doi.org/}%
\providecommand \selectlanguage [0]{\@gobble}%
\providecommand \bibinfo  [0]{\@secondoftwo}%
\providecommand \bibfield  [0]{\@secondoftwo}%
\providecommand \translation [1]{[#1]}%
\providecommand \BibitemOpen [0]{}%
\providecommand \bibitemStop [0]{}%
\providecommand \bibitemNoStop [0]{.\EOS\space}%
\providecommand \EOS [0]{\spacefactor3000\relax}%
\providecommand \BibitemShut  [1]{\csname bibitem#1\endcsname}%
\let\auto@bib@innerbib\@empty
\bibitem [{\citenamefont {Tittel}\ \emph {et~al.}(1998)\citenamefont {Tittel}, \citenamefont {Brendel}, \citenamefont {Zbinden},\ and\ \citenamefont {Gisin}}]{cod_1-PhysRevLett.81.3563}%
  \BibitemOpen
  \bibfield  {author} {\bibinfo {author} {\bibfnamefont {W.}~\bibnamefont {Tittel}}, \bibinfo {author} {\bibfnamefont {J.}~\bibnamefont {Brendel}}, \bibinfo {author} {\bibfnamefont {H.}~\bibnamefont {Zbinden}},\ and\ \bibinfo {author} {\bibfnamefont {N.}~\bibnamefont {Gisin}},\ }\bibfield  {title} {\bibinfo {title} {Violation of bell inequalities by photons more than 10 km apart},\ }\href {https://doi.org/10.1103/PhysRevLett.81.3563} {\bibfield  {journal} {\bibinfo  {journal} {Phys. Rev. Lett.}\ }\textbf {\bibinfo {volume} {81}},\ \bibinfo {pages} {3563} (\bibinfo {year} {1998})}\BibitemShut {NoStop}%
\bibitem [{\citenamefont {Mower}\ \emph {et~al.}(2013)\citenamefont {Mower}, \citenamefont {Zhang}, \citenamefont {Desjardins}, \citenamefont {Lee}, \citenamefont {Shapiro},\ and\ \citenamefont {Englund}}]{cod_2-PhysRevA.87.062322}%
  \BibitemOpen
  \bibfield  {author} {\bibinfo {author} {\bibfnamefont {J.}~\bibnamefont {Mower}}, \bibinfo {author} {\bibfnamefont {Z.}~\bibnamefont {Zhang}}, \bibinfo {author} {\bibfnamefont {P.}~\bibnamefont {Desjardins}}, \bibinfo {author} {\bibfnamefont {C.}~\bibnamefont {Lee}}, \bibinfo {author} {\bibfnamefont {J.~H.}\ \bibnamefont {Shapiro}},\ and\ \bibinfo {author} {\bibfnamefont {D.}~\bibnamefont {Englund}},\ }\bibfield  {title} {\bibinfo {title} {High-dimensional quantum key distribution using dispersive optics},\ }\href {https://doi.org/10.1103/PhysRevA.87.062322} {\bibfield  {journal} {\bibinfo  {journal} {Phys. Rev. A}\ }\textbf {\bibinfo {volume} {87}},\ \bibinfo {pages} {062322} (\bibinfo {year} {2013})}\BibitemShut {NoStop}%
\bibitem [{\citenamefont {Nunn}\ \emph {et~al.}(2013)\citenamefont {Nunn}, \citenamefont {Wright}, \citenamefont {S\"{o}ller}, \citenamefont {Zhang}, \citenamefont {Walmsley},\ and\ \citenamefont {Smith}}]{cod_3-capability3-Nunn:13}%
  \BibitemOpen
  \bibfield  {author} {\bibinfo {author} {\bibfnamefont {J.}~\bibnamefont {Nunn}}, \bibinfo {author} {\bibfnamefont {L.~J.}\ \bibnamefont {Wright}}, \bibinfo {author} {\bibfnamefont {C.}~\bibnamefont {S\"{o}ller}}, \bibinfo {author} {\bibfnamefont {L.}~\bibnamefont {Zhang}}, \bibinfo {author} {\bibfnamefont {I.~A.}\ \bibnamefont {Walmsley}},\ and\ \bibinfo {author} {\bibfnamefont {B.~J.}\ \bibnamefont {Smith}},\ }\bibfield  {title} {\bibinfo {title} {Large-alphabet time-frequency entangled quantum key distribution by means of time-to-frequency conversion},\ }\href {https://doi.org/10.1364/OE.21.015959} {\bibfield  {journal} {\bibinfo  {journal} {Opt. Express}\ }\textbf {\bibinfo {volume} {21}},\ \bibinfo {pages} {15959} (\bibinfo {year} {2013})}\BibitemShut {NoStop}%
\bibitem [{\citenamefont {Lukens}\ \emph {et~al.}(2014)\citenamefont {Lukens}, \citenamefont {Dezfooliyan}, \citenamefont {Langrock}, \citenamefont {Fejer}, \citenamefont {Leaird},\ and\ \citenamefont {Weiner}}]{cod_4-PhysRevLett.112.133602}%
  \BibitemOpen
  \bibfield  {author} {\bibinfo {author} {\bibfnamefont {J.~M.}\ \bibnamefont {Lukens}}, \bibinfo {author} {\bibfnamefont {A.}~\bibnamefont {Dezfooliyan}}, \bibinfo {author} {\bibfnamefont {C.}~\bibnamefont {Langrock}}, \bibinfo {author} {\bibfnamefont {M.~M.}\ \bibnamefont {Fejer}}, \bibinfo {author} {\bibfnamefont {D.~E.}\ \bibnamefont {Leaird}},\ and\ \bibinfo {author} {\bibfnamefont {A.~M.}\ \bibnamefont {Weiner}},\ }\bibfield  {title} {\bibinfo {title} {Orthogonal spectral coding of entangled photons},\ }\href {https://doi.org/10.1103/PhysRevLett.112.133602} {\bibfield  {journal} {\bibinfo  {journal} {Phys. Rev. Lett.}\ }\textbf {\bibinfo {volume} {112}},\ \bibinfo {pages} {133602} (\bibinfo {year} {2014})}\BibitemShut {NoStop}%
\bibitem [{\citenamefont {Brecht}\ \emph {et~al.}(2015)\citenamefont {Brecht}, \citenamefont {Reddy}, \citenamefont {Silberhorn},\ and\ \citenamefont {Raymer}}]{cod_5-PhysRevX.5.041017}%
  \BibitemOpen
  \bibfield  {author} {\bibinfo {author} {\bibfnamefont {B.}~\bibnamefont {Brecht}}, \bibinfo {author} {\bibfnamefont {D.~V.}\ \bibnamefont {Reddy}}, \bibinfo {author} {\bibfnamefont {C.}~\bibnamefont {Silberhorn}},\ and\ \bibinfo {author} {\bibfnamefont {M.~G.}\ \bibnamefont {Raymer}},\ }\bibfield  {title} {\bibinfo {title} {Photon temporal modes: A complete framework for quantum information science},\ }\href {https://doi.org/10.1103/PhysRevX.5.041017} {\bibfield  {journal} {\bibinfo  {journal} {Phys. Rev. X}\ }\textbf {\bibinfo {volume} {5}},\ \bibinfo {pages} {041017} (\bibinfo {year} {2015})}\BibitemShut {NoStop}%
\bibitem [{\citenamefont {Schwarz}\ \emph {et~al.}(2016)\citenamefont {Schwarz}, \citenamefont {Bessire}, \citenamefont {Stefanov},\ and\ \citenamefont {Liang}}]{cod_6-Schwarz_2016}%
  \BibitemOpen
  \bibfield  {author} {\bibinfo {author} {\bibfnamefont {S.}~\bibnamefont {Schwarz}}, \bibinfo {author} {\bibfnamefont {B.}~\bibnamefont {Bessire}}, \bibinfo {author} {\bibfnamefont {A.}~\bibnamefont {Stefanov}},\ and\ \bibinfo {author} {\bibfnamefont {Y.-C.}\ \bibnamefont {Liang}},\ }\bibfield  {title} {\bibinfo {title} {Bipartite bell inequalities with three ternary-outcome measurements—from theory to experiments},\ }\href {https://doi.org/10.1088/1367-2630/18/3/035001} {\bibfield  {journal} {\bibinfo  {journal} {New J. Phys.}\ }\textbf {\bibinfo {volume} {18}},\ \bibinfo {pages} {035001} (\bibinfo {year} {2016})}\BibitemShut {NoStop}%
\bibitem [{\citenamefont {Thiel}\ \emph {et~al.}(2017)\citenamefont {Thiel}, \citenamefont {Roslund}, \citenamefont {Jian}, \citenamefont {Fabre},\ and\ \citenamefont {Treps}}]{highdimension_Thiel_2017}%
  \BibitemOpen
  \bibfield  {author} {\bibinfo {author} {\bibfnamefont {V.}~\bibnamefont {Thiel}}, \bibinfo {author} {\bibfnamefont {J.}~\bibnamefont {Roslund}}, \bibinfo {author} {\bibfnamefont {P.}~\bibnamefont {Jian}}, \bibinfo {author} {\bibfnamefont {C.}~\bibnamefont {Fabre}},\ and\ \bibinfo {author} {\bibfnamefont {N.}~\bibnamefont {Treps}},\ }\bibfield  {title} {\bibinfo {title} {Quantum-limited measurements of distance fluctuations with a multimode detector},\ }\href {https://doi.org/10.1088/2058-9565/aa77d3} {\bibfield  {journal} {\bibinfo  {journal} {Quantum Sci. Technol.}\ }\textbf {\bibinfo {volume} {2}},\ \bibinfo {pages} {034008} (\bibinfo {year} {2017})}\BibitemShut {NoStop}%
\bibitem [{\citenamefont {Zhong}\ \emph {et~al.}(2015)\citenamefont {Zhong}, \citenamefont {Zhou}, \citenamefont {Horansky}, \citenamefont {Lee}, \citenamefont {Verma}, \citenamefont {Lita}, \citenamefont {Restelli}, \citenamefont {Bienfang}, \citenamefont {Mirin}, \citenamefont {Gerrits}, \citenamefont {Nam}, \citenamefont {Marsili}, \citenamefont {Shaw}, \citenamefont {Zhang}, \citenamefont {Wang}, \citenamefont {Englund}, \citenamefont {Wornell}, \citenamefont {Shapiro},\ and\ \citenamefont {Wong}}]{capability2-Zhong_2015}%
  \BibitemOpen
  \bibfield  {author} {\bibinfo {author} {\bibfnamefont {T.}~\bibnamefont {Zhong}}, \bibinfo {author} {\bibfnamefont {H.}~\bibnamefont {Zhou}}, \bibinfo {author} {\bibfnamefont {R.~D.}\ \bibnamefont {Horansky}}, \bibinfo {author} {\bibfnamefont {C.}~\bibnamefont {Lee}}, \bibinfo {author} {\bibfnamefont {V.~B.}\ \bibnamefont {Verma}}, \bibinfo {author} {\bibfnamefont {A.~E.}\ \bibnamefont {Lita}}, \bibinfo {author} {\bibfnamefont {A.}~\bibnamefont {Restelli}}, \bibinfo {author} {\bibfnamefont {J.~C.}\ \bibnamefont {Bienfang}}, \bibinfo {author} {\bibfnamefont {R.~P.}\ \bibnamefont {Mirin}}, \bibinfo {author} {\bibfnamefont {T.}~\bibnamefont {Gerrits}}, \bibinfo {author} {\bibfnamefont {S.~W.}\ \bibnamefont {Nam}}, \bibinfo {author} {\bibfnamefont {F.}~\bibnamefont {Marsili}}, \bibinfo {author} {\bibfnamefont {M.~D.}\ \bibnamefont {Shaw}}, \bibinfo {author} {\bibfnamefont {Z.}~\bibnamefont {Zhang}}, \bibinfo {author} {\bibfnamefont {L.}~\bibnamefont {Wang}}, \bibinfo {author} {\bibfnamefont {D.}~\bibnamefont
  {Englund}}, \bibinfo {author} {\bibfnamefont {G.~W.}\ \bibnamefont {Wornell}}, \bibinfo {author} {\bibfnamefont {J.~H.}\ \bibnamefont {Shapiro}},\ and\ \bibinfo {author} {\bibfnamefont {F.~N.~C.}\ \bibnamefont {Wong}},\ }\bibfield  {title} {\bibinfo {title} {Photon-efficient quantum key distribution using time–energy entanglement with high-dimensional encoding},\ }\href {https://doi.org/10.1088/1367-2630/17/2/022002} {\bibfield  {journal} {\bibinfo  {journal} {New J. Phys.}\ }\textbf {\bibinfo {volume} {17}},\ \bibinfo {pages} {022002} (\bibinfo {year} {2015})}\BibitemShut {NoStop}%
\bibitem [{\citenamefont {Smith}\ \emph {et~al.}(2009)\citenamefont {Smith}, \citenamefont {Mahou}, \citenamefont {Cohen}, \citenamefont {Lundeen},\ and\ \citenamefont {Walmsley}}]{Smith2009PhotonPG_MONO_jsi}%
  \BibitemOpen
  \bibfield  {author} {\bibinfo {author} {\bibfnamefont {B.~J.}\ \bibnamefont {Smith}}, \bibinfo {author} {\bibfnamefont {P.}~\bibnamefont {Mahou}}, \bibinfo {author} {\bibfnamefont {O.}~\bibnamefont {Cohen}}, \bibinfo {author} {\bibfnamefont {J.~S.}\ \bibnamefont {Lundeen}},\ and\ \bibinfo {author} {\bibfnamefont {I.~A.}\ \bibnamefont {Walmsley}},\ }\bibfield  {title} {\bibinfo {title} {Photon pair generation in birefringent optical fibers.},\ }\href {https://doi.org/10.1364/oe.17.023589} {\bibfield  {journal} {\bibinfo  {journal} {Opt. Express}\ }\textbf {\bibinfo {volume} {17 26}},\ \bibinfo {pages} {23589} (\bibinfo {year} {2009})}\BibitemShut {NoStop}%
\bibitem [{\citenamefont {Davis}\ \emph {et~al.}(2017)\citenamefont {Davis}, \citenamefont {Saulnier}, \citenamefont {Karpi{\'n}ski},\ and\ \citenamefont {Smith}}]{Davis2016PulsedSS_CHIRPED_JSI}%
  \BibitemOpen
  \bibfield  {author} {\bibinfo {author} {\bibfnamefont {A.~O.}\ \bibnamefont {Davis}}, \bibinfo {author} {\bibfnamefont {P.~M.}\ \bibnamefont {Saulnier}}, \bibinfo {author} {\bibfnamefont {M.}~\bibnamefont {Karpi{\'n}ski}},\ and\ \bibinfo {author} {\bibfnamefont {B.~J.}\ \bibnamefont {Smith}},\ }\bibfield  {title} {\bibinfo {title} {Pulsed single-photon spectrometer by frequency-to-time mapping using chirped fiber bragg gratings},\ }\href {https://doi.org/10.1364/oe.25.012804} {\bibfield  {journal} {\bibinfo  {journal} {Optics Express}\ }\textbf {\bibinfo {volume} {25}},\ \bibinfo {pages} {12804} (\bibinfo {year} {2017})}\BibitemShut {NoStop}%
\bibitem [{\citenamefont {Trebino}\ and\ \citenamefont {Kane}(1993)}]{Trebino1993UsingPR_Trebino}%
  \BibitemOpen
  \bibfield  {author} {\bibinfo {author} {\bibfnamefont {R.}~\bibnamefont {Trebino}}\ and\ \bibinfo {author} {\bibfnamefont {D.~J.}\ \bibnamefont {Kane}},\ }\bibfield  {title} {\bibinfo {title} {Using phase retrieval to measure the intensity and phase of ultrashort pulses: frequency-resolved optical gating},\ }\href {https://doi.org/10.1364/JOSAA.10.001101} {\bibfield  {journal} {\bibinfo  {journal} {J. Opt. Soc. Am. A}\ }\textbf {\bibinfo {volume} {10}},\ \bibinfo {pages} {1101} (\bibinfo {year} {1993})}\BibitemShut {NoStop}%
\bibitem [{\citenamefont {Iaconis}\ and\ \citenamefont {Walmsley}(1998)}]{SPIDER-Iaconis:98}%
  \BibitemOpen
  \bibfield  {author} {\bibinfo {author} {\bibfnamefont {C.}~\bibnamefont {Iaconis}}\ and\ \bibinfo {author} {\bibfnamefont {I.~A.}\ \bibnamefont {Walmsley}},\ }\bibfield  {title} {\bibinfo {title} {Spectral phase interferometry for direct electric-field reconstruction of ultrashort optical pulses},\ }\href {https://doi.org/10.1364/OL.23.000792} {\bibfield  {journal} {\bibinfo  {journal} {Opt. Lett.}\ }\textbf {\bibinfo {volume} {23}},\ \bibinfo {pages} {792} (\bibinfo {year} {1998})}\BibitemShut {NoStop}%
\bibitem [{\citenamefont {Davis}\ \emph {et~al.}(2018)\citenamefont {Davis}, \citenamefont {Thiel}, \citenamefont {Karpi\ifmmode~\acute{n}\else \'{n}\fi{}ski},\ and\ \citenamefont {Smith}}]{PhysRevLett.121.083602_shearing_singlephoton}%
  \BibitemOpen
  \bibfield  {author} {\bibinfo {author} {\bibfnamefont {A.~O.~C.}\ \bibnamefont {Davis}}, \bibinfo {author} {\bibfnamefont {V.}~\bibnamefont {Thiel}}, \bibinfo {author} {\bibfnamefont {M.}~\bibnamefont {Karpi\ifmmode~\acute{n}\else \'{n}\fi{}ski}},\ and\ \bibinfo {author} {\bibfnamefont {B.~J.}\ \bibnamefont {Smith}},\ }\bibfield  {title} {\bibinfo {title} {Measuring the single-photon temporal-spectral wave function},\ }\href {https://doi.org/10.1103/PhysRevLett.121.083602} {\bibfield  {journal} {\bibinfo  {journal} {Phys. Rev. Lett.}\ }\textbf {\bibinfo {volume} {121}},\ \bibinfo {pages} {083602} (\bibinfo {year} {2018})}\BibitemShut {NoStop}%
\bibitem [{\citenamefont {Ansari}\ \emph {et~al.}(2018)\citenamefont {Ansari}, \citenamefont {Donohue}, \citenamefont {Allgaier}, \citenamefont {Sansoni}, \citenamefont {Brecht}, \citenamefont {Roslund}, \citenamefont {Treps}, \citenamefont {Harder},\ and\ \citenamefont {Silberhorn}}]{ansari2018tomography_DAVIS23}%
  \BibitemOpen
  \bibfield  {author} {\bibinfo {author} {\bibfnamefont {V.}~\bibnamefont {Ansari}}, \bibinfo {author} {\bibfnamefont {J.~M.}\ \bibnamefont {Donohue}}, \bibinfo {author} {\bibfnamefont {M.}~\bibnamefont {Allgaier}}, \bibinfo {author} {\bibfnamefont {L.}~\bibnamefont {Sansoni}}, \bibinfo {author} {\bibfnamefont {B.}~\bibnamefont {Brecht}}, \bibinfo {author} {\bibfnamefont {J.}~\bibnamefont {Roslund}}, \bibinfo {author} {\bibfnamefont {N.}~\bibnamefont {Treps}}, \bibinfo {author} {\bibfnamefont {G.}~\bibnamefont {Harder}},\ and\ \bibinfo {author} {\bibfnamefont {C.}~\bibnamefont {Silberhorn}},\ }\bibfield  {title} {\bibinfo {title} {Tomography and purification of the temporal-mode structure of quantum light},\ }\href {https://doi.org/10.1103/PhysRevLett.120.213601} {\bibfield  {journal} {\bibinfo  {journal} {Phys. Rev. Lett.}\ }\textbf {\bibinfo {volume} {120}},\ \bibinfo {pages} {213601} (\bibinfo {year} {2018})}\BibitemShut {NoStop}%
\bibitem [{\citenamefont {Huo}\ \emph {et~al.}(2020)\citenamefont {Huo}, \citenamefont {Liu}, \citenamefont {Li}, \citenamefont {Cui}, \citenamefont {Chen}, \citenamefont {Palivela}, \citenamefont {Xie}, \citenamefont {Li},\ and\ \citenamefont {Ou}}]{PhysRevLett.124.213603_Nanshijie}%
  \BibitemOpen
  \bibfield  {author} {\bibinfo {author} {\bibfnamefont {N.}~\bibnamefont {Huo}}, \bibinfo {author} {\bibfnamefont {Y.}~\bibnamefont {Liu}}, \bibinfo {author} {\bibfnamefont {J.}~\bibnamefont {Li}}, \bibinfo {author} {\bibfnamefont {L.}~\bibnamefont {Cui}}, \bibinfo {author} {\bibfnamefont {X.}~\bibnamefont {Chen}}, \bibinfo {author} {\bibfnamefont {R.}~\bibnamefont {Palivela}}, \bibinfo {author} {\bibfnamefont {T.}~\bibnamefont {Xie}}, \bibinfo {author} {\bibfnamefont {X.}~\bibnamefont {Li}},\ and\ \bibinfo {author} {\bibfnamefont {Z.~Y.}\ \bibnamefont {Ou}},\ }\bibfield  {title} {\bibinfo {title} {Direct temporal mode measurement for the characterization of temporally multiplexed high dimensional quantum entanglement in continuous variables},\ }\href {https://doi.org/10.1103/PhysRevLett.124.213603} {\bibfield  {journal} {\bibinfo  {journal} {Phys. Rev. Lett.}\ }\textbf {\bibinfo {volume} {124}},\ \bibinfo {pages} {213603} (\bibinfo {year} {2020})}\BibitemShut {NoStop}%
\bibitem [{\citenamefont {Wasilewski}\ \emph {et~al.}(2007)\citenamefont {Wasilewski}, \citenamefont {Kolenderski},\ and\ \citenamefont {Frankowski}}]{wasilewski2007spectral_DAVIS20}%
  \BibitemOpen
  \bibfield  {author} {\bibinfo {author} {\bibfnamefont {W.}~\bibnamefont {Wasilewski}}, \bibinfo {author} {\bibfnamefont {P.}~\bibnamefont {Kolenderski}},\ and\ \bibinfo {author} {\bibfnamefont {R.}~\bibnamefont {Frankowski}},\ }\bibfield  {title} {\bibinfo {title} {Spectral density matrix of a single photon measured},\ }\href {https://doi.org/10.1103/physrevlett.99.123601} {\bibfield  {journal} {\bibinfo  {journal} {Phys. Rev. Lett.}\ }\textbf {\bibinfo {volume} {99}},\ \bibinfo {pages} {123601} (\bibinfo {year} {2007})}\BibitemShut {NoStop}%
\bibitem [{\citenamefont {Thiel}\ \emph {et~al.}(2019)\citenamefont {Thiel}, \citenamefont {Davis}, \citenamefont {Sun}, \citenamefont {D'Ornellas}, \citenamefont {Jin},\ and\ \citenamefont {Smith}}]{Thiel:19_FrqRslvHOM_singlephoton}%
  \BibitemOpen
  \bibfield  {author} {\bibinfo {author} {\bibfnamefont {V.}~\bibnamefont {Thiel}}, \bibinfo {author} {\bibfnamefont {A.~O.~C.}\ \bibnamefont {Davis}}, \bibinfo {author} {\bibfnamefont {K.}~\bibnamefont {Sun}}, \bibinfo {author} {\bibfnamefont {P.}~\bibnamefont {D'Ornellas}}, \bibinfo {author} {\bibfnamefont {X.-M.}\ \bibnamefont {Jin}},\ and\ \bibinfo {author} {\bibfnamefont {B.~J.}\ \bibnamefont {Smith}},\ }\bibfield  {title} {\bibinfo {title} {Single-photon characterization by spectrally-resolved hong-ou-mandel interference},\ }in\ \href {https://doi.org/10.1364/CQO.2019.M5A.21} {\emph {\bibinfo {booktitle} {Rochester Conference on Coherence and Quantum Optics (CQO-11)}}}\ (\bibinfo  {publisher} {Optica Publishing Group},\ \bibinfo {year} {2019})\ p.\ \bibinfo {pages} {M5A.21}\BibitemShut {NoStop}%
\bibitem [{\citenamefont {Qin}\ \emph {et~al.}(2015)\citenamefont {Qin}, \citenamefont {Prasad}, \citenamefont {Brannan}, \citenamefont {MacRae}, \citenamefont {Lezama},\ and\ \citenamefont {Lvovsky}}]{qin2015complete_DAVIS22}%
  \BibitemOpen
  \bibfield  {author} {\bibinfo {author} {\bibfnamefont {Z.}~\bibnamefont {Qin}}, \bibinfo {author} {\bibfnamefont {A.~S.}\ \bibnamefont {Prasad}}, \bibinfo {author} {\bibfnamefont {T.}~\bibnamefont {Brannan}}, \bibinfo {author} {\bibfnamefont {A.}~\bibnamefont {MacRae}}, \bibinfo {author} {\bibfnamefont {A.}~\bibnamefont {Lezama}},\ and\ \bibinfo {author} {\bibfnamefont {A.}~\bibnamefont {Lvovsky}},\ }\bibfield  {title} {\bibinfo {title} {Complete temporal characterization of a single photon},\ }\href {https://doi.org/10.1038/lsa.2015.71} {\bibfield  {journal} {\bibinfo  {journal} {Light Sci. Appl.}\ }\textbf {\bibinfo {volume} {4}},\ \bibinfo {pages} {e298} (\bibinfo {year} {2015})}\BibitemShut {NoStop}%
\bibitem [{\citenamefont {Polycarpou}\ \emph {et~al.}(2012)\citenamefont {Polycarpou}, \citenamefont {Cassemiro}, \citenamefont {Venturi}, \citenamefont {Zavatta},\ and\ \citenamefont {Bellini}}]{polycarpou2012adaptive_DAVIS21}%
  \BibitemOpen
  \bibfield  {author} {\bibinfo {author} {\bibfnamefont {C.}~\bibnamefont {Polycarpou}}, \bibinfo {author} {\bibfnamefont {K.~N.}\ \bibnamefont {Cassemiro}}, \bibinfo {author} {\bibfnamefont {G.}~\bibnamefont {Venturi}}, \bibinfo {author} {\bibfnamefont {A.}~\bibnamefont {Zavatta}},\ and\ \bibinfo {author} {\bibfnamefont {M.}~\bibnamefont {Bellini}},\ }\bibfield  {title} {\bibinfo {title} {Adaptive detection of arbitrarily shaped ultrashort quantum light states},\ }\href {https://doi.org/10.1103/PHYSREVLETT.109.053602} {\bibfield  {journal} {\bibinfo  {journal} {Phys. Rev. Lett.}\ }\textbf {\bibinfo {volume} {109}},\ \bibinfo {pages} {053602} (\bibinfo {year} {2012})}\BibitemShut {NoStop}%
\bibitem [{\citenamefont {Hong}\ \emph {et~al.}(1987)\citenamefont {Hong}, \citenamefont {Ou},\ and\ \citenamefont {Mandel}}]{PhysRevLett.59.2044}%
  \BibitemOpen
  \bibfield  {author} {\bibinfo {author} {\bibfnamefont {C.~K.}\ \bibnamefont {Hong}}, \bibinfo {author} {\bibfnamefont {Z.~Y.}\ \bibnamefont {Ou}},\ and\ \bibinfo {author} {\bibfnamefont {L.}~\bibnamefont {Mandel}},\ }\bibfield  {title} {\bibinfo {title} {Measurement of subpicosecond time intervals between two photons by interference},\ }\href {https://doi.org/10.1103/physrevlett.59.2044} {\bibfield  {journal} {\bibinfo  {journal} {Phys. Rev. Lett.}\ }\textbf {\bibinfo {volume} {59}},\ \bibinfo {pages} {2044—2046} (\bibinfo {year} {1987})}\BibitemShut {NoStop}%
\bibitem [{\citenamefont {Menssen}\ \emph {et~al.}(2017)\citenamefont {Menssen}, \citenamefont {Jones}, \citenamefont {Metcalf}, \citenamefont {Tichy}, \citenamefont {Barz}, \citenamefont {Kolthammer},\ and\ \citenamefont {Walmsley}}]{PhysRevLett.118.153603_Disting}%
  \BibitemOpen
  \bibfield  {author} {\bibinfo {author} {\bibfnamefont {A.~J.}\ \bibnamefont {Menssen}}, \bibinfo {author} {\bibfnamefont {A.~E.}\ \bibnamefont {Jones}}, \bibinfo {author} {\bibfnamefont {B.~J.}\ \bibnamefont {Metcalf}}, \bibinfo {author} {\bibfnamefont {M.~C.}\ \bibnamefont {Tichy}}, \bibinfo {author} {\bibfnamefont {S.}~\bibnamefont {Barz}}, \bibinfo {author} {\bibfnamefont {W.~S.}\ \bibnamefont {Kolthammer}},\ and\ \bibinfo {author} {\bibfnamefont {I.~A.}\ \bibnamefont {Walmsley}},\ }\bibfield  {title} {\bibinfo {title} {Distinguishability and many-particle interference},\ }\href {https://doi.org/10.1103/PhysRevLett.118.153603} {\bibfield  {journal} {\bibinfo  {journal} {Phys. Rev. Lett.}\ }\textbf {\bibinfo {volume} {118}},\ \bibinfo {pages} {153603} (\bibinfo {year} {2017})}\BibitemShut {NoStop}%
\bibitem [{\citenamefont {Ou}\ and\ \citenamefont {Li}(2022)}]{PhysRevResearch.4.023125_LI_OU_FOUR_ORDER}%
  \BibitemOpen
  \bibfield  {author} {\bibinfo {author} {\bibfnamefont {Z.~Y.}\ \bibnamefont {Ou}}\ and\ \bibinfo {author} {\bibfnamefont {X.}~\bibnamefont {Li}},\ }\bibfield  {title} {\bibinfo {title} {Unbalanced fourth-order interference beyond coherence time},\ }\href {https://doi.org/10.1103/PhysRevResearch.4.023125} {\bibfield  {journal} {\bibinfo  {journal} {Phys. Rev. Res.}\ }\textbf {\bibinfo {volume} {4}},\ \bibinfo {pages} {023125} (\bibinfo {year} {2022})}\BibitemShut {NoStop}%
\bibitem [{\citenamefont {Ma}\ \emph {et~al.}(2015)\citenamefont {Ma}, \citenamefont {Cui},\ and\ \citenamefont {Li}}]{Xiaoxin2015Hong}%
  \BibitemOpen
  \bibfield  {author} {\bibinfo {author} {\bibfnamefont {X.}~\bibnamefont {Ma}}, \bibinfo {author} {\bibfnamefont {L.}~\bibnamefont {Cui}},\ and\ \bibinfo {author} {\bibfnamefont {X.}~\bibnamefont {Li}},\ }\bibfield  {title} {\bibinfo {title} {Hong–ou–mandel interference between independent sources of heralded ultrafast single photons: influence of chirp},\ }\href {https://opg.optica.org/josab/abstract.cfm?URI=josab-32-5-946} {\bibfield  {journal} {\bibinfo  {journal} {J. Opt. Soc. Am. B}\ }\textbf {\bibinfo {volume} {32}},\ \bibinfo {pages} {946} (\bibinfo {year} {2015})}\BibitemShut {NoStop}%
\bibitem [{\citenamefont {Ma}\ \emph {et~al.}(2011)\citenamefont {Ma}, \citenamefont {Li}, \citenamefont {Cui}, \citenamefont {Guo},\ and\ \citenamefont {Yang}}]{ma2011effect}%
  \BibitemOpen
  \bibfield  {author} {\bibinfo {author} {\bibfnamefont {X.}~\bibnamefont {Ma}}, \bibinfo {author} {\bibfnamefont {X.}~\bibnamefont {Li}}, \bibinfo {author} {\bibfnamefont {L.}~\bibnamefont {Cui}}, \bibinfo {author} {\bibfnamefont {X.}~\bibnamefont {Guo}},\ and\ \bibinfo {author} {\bibfnamefont {L.}~\bibnamefont {Yang}},\ }\bibfield  {title} {\bibinfo {title} {Effect of chromatic-dispersion-induced chirp on the temporal coherence properties of individual beams from spontaneous four-wave mixing},\ }\href {https://doi.org/10.1103/PhysRevA.84.023829} {\bibfield  {journal} {\bibinfo  {journal} {Phys. Rev. A}\ }\textbf {\bibinfo {volume} {84}},\ \bibinfo {pages} {023829} (\bibinfo {year} {2011})}\BibitemShut {NoStop}%
\bibitem [{\citenamefont {Fan}\ \emph {et~al.}(2021)\citenamefont {Fan}, \citenamefont {Yuan}, \citenamefont {Zhang}, \citenamefont {Shen}, \citenamefont {Wu}, \citenamefont {Wang}, \citenamefont {Li}, \citenamefont {Deng}, \citenamefont {Song}, \citenamefont {You}, \citenamefont {Wang}, \citenamefont {Wang}, \citenamefont {Guo},\ and\ \citenamefont {Zhou}}]{fan2021effect}%
  \BibitemOpen
  \bibfield  {author} {\bibinfo {author} {\bibfnamefont {Y.-R.}\ \bibnamefont {Fan}}, \bibinfo {author} {\bibfnamefont {C.-Z.}\ \bibnamefont {Yuan}}, \bibinfo {author} {\bibfnamefont {R.-M.}\ \bibnamefont {Zhang}}, \bibinfo {author} {\bibfnamefont {S.}~\bibnamefont {Shen}}, \bibinfo {author} {\bibfnamefont {P.}~\bibnamefont {Wu}}, \bibinfo {author} {\bibfnamefont {H.-Q.}\ \bibnamefont {Wang}}, \bibinfo {author} {\bibfnamefont {H.}~\bibnamefont {Li}}, \bibinfo {author} {\bibfnamefont {G.-W.}\ \bibnamefont {Deng}}, \bibinfo {author} {\bibfnamefont {H.-Z.}\ \bibnamefont {Song}}, \bibinfo {author} {\bibfnamefont {L.-X.}\ \bibnamefont {You}}, \bibinfo {author} {\bibfnamefont {Z.}~\bibnamefont {Wang}}, \bibinfo {author} {\bibfnamefont {Y.}~\bibnamefont {Wang}}, \bibinfo {author} {\bibfnamefont {G.-C.}\ \bibnamefont {Guo}},\ and\ \bibinfo {author} {\bibfnamefont {Q.}~\bibnamefont {Zhou}},\ }\bibfield  {title} {\bibinfo {title} {Effect of dispersion on indistinguishability between single-photon wave-packets},\ }\href
  {https://doi.org/10.1364/PRJ.421180} {\bibfield  {journal} {\bibinfo  {journal} {Photon. Res.}\ }\textbf {\bibinfo {volume} {9}},\ \bibinfo {pages} {1134} (\bibinfo {year} {2021})}\BibitemShut {NoStop}%
\bibitem [{\citenamefont {Miyamoto}\ \emph {et~al.}(1993)\citenamefont {Miyamoto}, \citenamefont {Kuga}, \citenamefont {Baba},\ and\ \citenamefont {Matsuoka}}]{Y1993Measurement}%
  \BibitemOpen
  \bibfield  {author} {\bibinfo {author} {\bibfnamefont {Y.}~\bibnamefont {Miyamoto}}, \bibinfo {author} {\bibfnamefont {T.}~\bibnamefont {Kuga}}, \bibinfo {author} {\bibfnamefont {M.}~\bibnamefont {Baba}},\ and\ \bibinfo {author} {\bibfnamefont {M.}~\bibnamefont {Matsuoka}},\ }\bibfield  {title} {\bibinfo {title} {Measurement of ultrafast optical pulses with two-photon interference},\ }\href {https://doi.org/10.1364/OL.18.000900} {\bibfield  {journal} {\bibinfo  {journal} {Opt. Lett.}\ }\textbf {\bibinfo {volume} {18}},\ \bibinfo {pages} {900} (\bibinfo {year} {1993})}\BibitemShut {NoStop}%
\bibitem [{\citenamefont {Zhao}\ \emph {et~al.}(2022)\citenamefont {Zhao}, \citenamefont {Huo}, \citenamefont {Cui}, \citenamefont {Li},\ and\ \citenamefont {Ou}}]{2021Propagation}%
  \BibitemOpen
  \bibfield  {author} {\bibinfo {author} {\bibfnamefont {W.}~\bibnamefont {Zhao}}, \bibinfo {author} {\bibfnamefont {N.}~\bibnamefont {Huo}}, \bibinfo {author} {\bibfnamefont {L.}~\bibnamefont {Cui}}, \bibinfo {author} {\bibfnamefont {X.}~\bibnamefont {Li}},\ and\ \bibinfo {author} {\bibfnamefont {Z.~Y.}\ \bibnamefont {Ou}},\ }\bibfield  {title} {\bibinfo {title} {Propagation of temporal mode multiplexed optical fields in fibers: influence of dispersion},\ }\href {https://doi.org/10.1364/OE.448013} {\bibfield  {journal} {\bibinfo  {journal} {Opt. Express}\ }\textbf {\bibinfo {volume} {30}},\ \bibinfo {pages} {447} (\bibinfo {year} {2022})}\BibitemShut {NoStop}%
\bibitem [{\citenamefont {Fienup}(1982)}]{J1982Phase}%
  \BibitemOpen
  \bibfield  {author} {\bibinfo {author} {\bibfnamefont {J.~R.}\ \bibnamefont {Fienup}},\ }\bibfield  {title} {\bibinfo {title} {Phase retrieval algorithms: a comparison},\ }\href {https://api.semanticscholar.org/CorpusID:10777701} {\bibfield  {journal} {\bibinfo  {journal} {Appl. Opt.}\ }\textbf {\bibinfo {volume} {21 15}},\ \bibinfo {pages} {2758} (\bibinfo {year} {1982})}\BibitemShut {NoStop}%
\bibitem [{\citenamefont {Gerchberg}(1972)}]{Gerchberg1972APA}%
  \BibitemOpen
  \bibfield  {author} {\bibinfo {author} {\bibfnamefont {R.~W.}\ \bibnamefont {Gerchberg}},\ }\bibfield  {title} {\bibinfo {title} {A practical algorithm for the determination of phase from image and diffraction plane pictures},\ }\href {https://api.semanticscholar.org/CorpusID:55691159} {\bibfield  {journal} {\bibinfo  {journal} {Optik}\ }\textbf {\bibinfo {volume} {35}},\ \bibinfo {pages} {237} (\bibinfo {year} {1972})}\BibitemShut {NoStop}%
\bibitem [{\citenamefont {Yudilevich}\ \emph {et~al.}(1987)\citenamefont {Yudilevich}, \citenamefont {Levi}, \citenamefont {Habetler},\ and\ \citenamefont {Stark}}]{Yudilevich1987RestorationOS_GPimage}%
  \BibitemOpen
  \bibfield  {author} {\bibinfo {author} {\bibfnamefont {E.}~\bibnamefont {Yudilevich}}, \bibinfo {author} {\bibfnamefont {A.}~\bibnamefont {Levi}}, \bibinfo {author} {\bibfnamefont {G.~J.}\ \bibnamefont {Habetler}},\ and\ \bibinfo {author} {\bibfnamefont {H.}~\bibnamefont {Stark}},\ }\bibfield  {title} {\bibinfo {title} {Restoration of signals from their signed fourier-transform magnitude by the method of generalized projections},\ }\href {https://doi.org/10.1364/JOSAA.4.000236} {\bibfield  {journal} {\bibinfo  {journal} {J. Opt. Soc. Am. A}\ }\textbf {\bibinfo {volume} {4}},\ \bibinfo {pages} {236} (\bibinfo {year} {1987})}\BibitemShut {NoStop}%
\bibitem [{\citenamefont {DeLong}\ \emph {et~al.}(1994)\citenamefont {DeLong}, \citenamefont {Trebino}, \citenamefont {Hunter},\ and\ \citenamefont {White}}]{1994Frequency}%
  \BibitemOpen
  \bibfield  {author} {\bibinfo {author} {\bibfnamefont {K.~W.}\ \bibnamefont {DeLong}}, \bibinfo {author} {\bibfnamefont {R.}~\bibnamefont {Trebino}}, \bibinfo {author} {\bibfnamefont {J.}~\bibnamefont {Hunter}},\ and\ \bibinfo {author} {\bibfnamefont {W.~E.}\ \bibnamefont {White}},\ }\bibfield  {title} {\bibinfo {title} {Frequency-resolved optical gating with the use of second-harmonic generation},\ }\href {https://doi.org/10.1364/JOSAB.11.002206} {\bibfield  {journal} {\bibinfo  {journal} {J. Opt. Soc. Am. B}\ }\textbf {\bibinfo {volume} {11}},\ \bibinfo {pages} {2206} (\bibinfo {year} {1994})}\BibitemShut {NoStop}%
\end{thebibliography}%

\end{document}